\documentclass[prb,reprint,preprintnumbers,
amsmath,amssymb, aps, floatfix
]{revtex4-2}

\usepackage{graphicx}
\usepackage{dcolumn}
\usepackage{bm}
\usepackage{amssymb}
\usepackage{amsmath}
\usepackage{color}
\usepackage{soul}
\usepackage{subfig}
\usepackage{appendix}
\usepackage{booktabs}
\usepackage{comment}
\usepackage{epstopdf}
\usepackage[hyperindex,breaklinks]{hyperref}

\begin{document}
	
	\title{Vortex-Driven Superconducting Diode Effect in Asymmetric Multilayer Heterostructures}
	
	\author{Jiong Li$^{1, 2}$}
	\author{Ji Jiang$^3$}
	\author{Qing-Hu Chen$^{2,4,}$}
	\email{qhchen@zju.edu.cn}
	
	\affiliation{
		$^1$ Institute for Advanced Study in Physics, Zhejiang University, Hangzhou 310027, China \\
		$^2$ Zhejiang Key Laboratory of Micro-Nano Quantum Chips and Quantum Control, School of Physics, Zhejiang University, Hangzhou 310027, China. \\
		$^3$ International Quantum Academy, Shenzhen 518048, China. \\
		$^4$ Collaborative Innovation Center of Advanced Microstructures, Nanjing University, Nanjing 210093, China.
	}
	
	\begin{abstract}
		The superconducting diode effect (SDE), characterized by nonreciprocal critical currents, has attracted growing attention due to its potential applications in quantum technologies and energy-efficient devices. In this work, we explore the microscopic mechanism of the SDE by simulating asymmetric multilayer heterostructures within time-dependent Ginzburg–Landau theory. We systematically vary the layer thickness, external magnetic field and stacking order in a trilayer structure composed of niobium, vanadium, and tantalum, which share a similar structure to that in the pioneering experimental work, to clarify the role of vortex dynamics. Our simulations reveal a pronounced SDE originating from the interplay of Lorentz forces and asymmetric vortex dynamics, which strongly depends on layer stacking order. Besides, by simply changing the stacking order of the constituent layers, the SDE can be entirely suppressed. These findings offer insights into the microscopic mechanisms of the SDE and provide a feasible approach for controlling and eliminating the SDE in practical superconducting devices.
	\end{abstract}
	
	\maketitle
	
	\section{Introduction}
	
	Directional transport and the propagation of physical quantities are well-established phenomena in materials lacking inversion symmetry \cite{tokura_nonreciprocal_2018}. This effect, extensively studied in semiconductors, plays a crucial role in a wide range of technologies, including the topological magnetoelectric effect \cite{avci_unidirectional_2015, morimoto_topological_2015, yasuda_large_2016}, polar semiconductors \cite{rikken_electrical_2001, kubota_x-ray_2004, kezsmarki_enhanced_2011}, and p-n junctions \cite{shockley_theory_1949, nakamura_high-power_1991, feng_all_2018}. A representative example is the semiconductor diode, which allows unidirectional current flow and serves as a cornerstone of modern computation, communication, and sensing technologies \cite{ideue_bulk_2017, rein_diode_2018, 2020One}.
	
	Nonreciprocity in superconductors, particularly the superconducting diode effect (SDE), has attracted considerable attention because it enables zero resistance for current flow in only one direction. This distinctive property holds transformative potential for quantum computing, high-sensitivity sensing, and energy transmission \cite{wakatsuki_nonreciprocal_2017, cui_transport_2019, itahashi_nonreciprocal_2020, bauriedl_supercurrent_2022, wang_anisotropic_2023, nadeem_superconducting_2023}. A groundbreaking study on an artificial superlattice, $[\mathrm{Nb/V/Ta}]_{n}$—which breaks spatial inversion symmetry through its asymmetric layer structure and violates time-reversal symmetry under an external magnetic field—first demonstrated the SDE \cite{ando_observation_2020}. These results were rapidly reproduced, triggering a surge of research interest \cite{miyasaka_observation_2021, kawarazaki_magnetic-field-induced_2022}. Since then, the SDE has been observed in a variety of superconducting systems, including polar superconducting films \cite{kealhofer_anomalous_2023, hou_ubiquitous_2023, du_superconducting_2024}, Josephson junctions \cite{wu_field-free_2022, pal_josephson_2022, trahms_diode_2023, kim_intrinsic_2024}, superconductor/ferromagnet bilayers \cite{narita_field-free_2022, karabassov_hybrid_2022, gutfreund_direct_2023}, and twisted trilayer graphene \cite{lin_zero-field_2022}. Considerable theoretical efforts have been devoted to explaining the SDE \cite{jiang_reversible_2021, misaki_theory_2021, daido_intrinsic_2022, ilic_theory_2022, he_phenomenological_2022, jiang_field-free_2022, zinkl_symmetry_2022, legg_superconducting_2022, daido_superconducting_2022, tanaka_theory_2022, he_supercurrent_2023, li_unconventional_2024, matsumoto_reciprocal_2025}, many of which are based on the Ginzburg–Landau (GL) theoretical framework \cite{daido_intrinsic_2022, he_phenomenological_2022, daido_superconducting_2022, zinkl_symmetry_2022, he_supercurrent_2023}. Despite the pioneering observation of the SDE in the artificial superlattice $\mathrm{[Nb(1.0\,nm)/V(1.0\,nm)/Ta(1.0\,nm)]}_{40}$ \cite{ando_observation_2020}, its microscopic origin remains incompletely understood.
	
	Fu~\cite{fu2021one} suggested  in a commentary that the orbital effect of a parallel magnetic field could significantly influence this SDE. Ando \textit{et al.}~\cite{ando_observation_2020} focused on the Rashba spin–orbit effect and its potential influence on unconventional pairing symmetry in the superconducting state. They also argue that the nonreciprocal charge transport and critical current are not fully understood until more accurate estimates of the pairing compositions in the [Nb/V/Ta]n superlattice are obtained. Vortices are likely to enter along the field direction because the coherence length is much smaller than the thickness of the superlattice. These authors acknowledge that the vortex dynamics was neglected in their analysis, which could be the other mechanism partially responsible for the observed SDE.  Ideue and Iwasa later suggested  ~\cite{2020One} that vortex dynamics may also contribute to the SDE. Theoretically, the role of vortices in the observed SDE cannot be neglected \cite{hou_ubiquitous_2023, moll_evolution_2023}, as vortex dynamics cause the experimentally observed critical current to be generally lower than the depairing critical current.	
	
	\begin{figure}[tbp]
		\centering
		\includegraphics[width=1.0\linewidth]{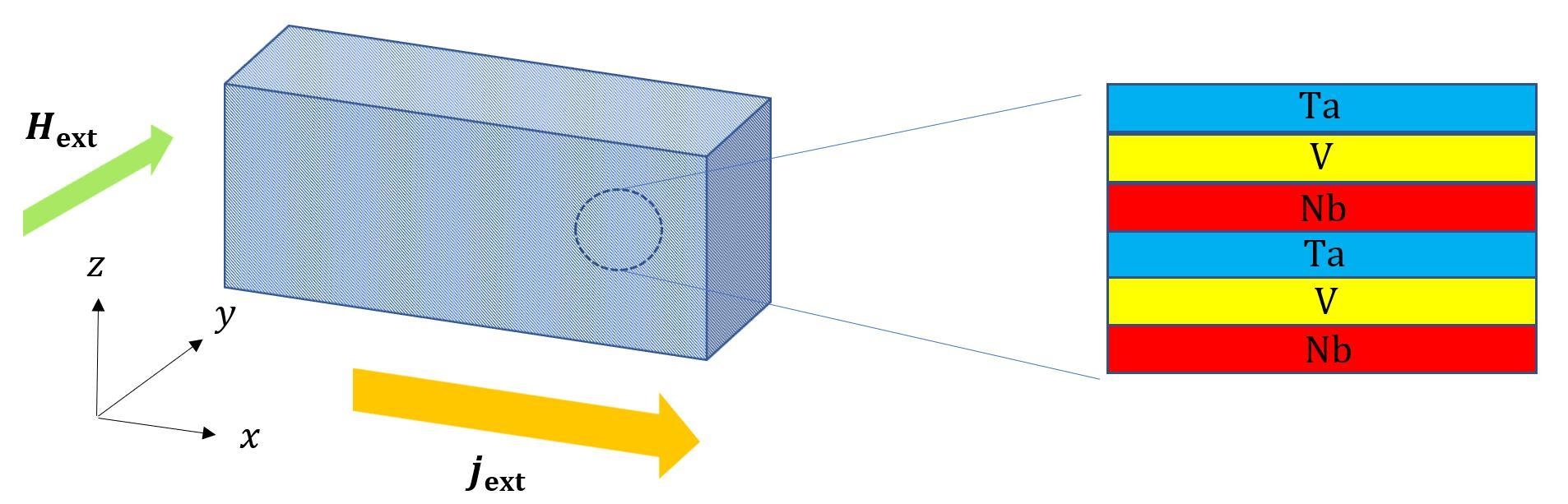}
		\caption{\textbf{Illustration of the multilayer superconducting structure.} A three-dimensional schematic of the superconducting sample, shown as a blue cuboid. A yellow arrow indicates the direction of the applied external current $\mathbf{j_\mathrm{ext}}$ along the $x$-axis, and a green arrow indicates the direction of the external magnetic field $\mathbf{H_\mathrm{ext}}$ along the $y$-axis. The coordinate axes are shown in black. The enlarged view of the periodic layered structure highlighted by the dashed circle is shown as color-coded, with red representing niobium, yellow denoting vanadium, and blue being tantalum.}
		\label{fig_model}
	\end{figure}
	
	To isolate the contribution of vortex dynamics to the SDE in a comparable non-centrosymmetric superlattice, we construct a simulation model by periodically stacking three types of superconducting layers in the form $[\mathrm{Nb/V/Ta}]_{n}$. Unlike the work of Ando \textit{et al.} \cite{ando_observation_2020}, we employ thicker layers to suppress the effects of Zeeman splitting and spin–orbit coupling. A schematic illustration of the sample is shown in Fig.~\ref{fig_model}. Our analysis focuses on vortex motion under an external magnetic field within the framework of time-dependent Ginzburg–Landau (TDGL) theory. The GL theory is widely regarded as an effective tool for describing the equilibrium macroscopic and mesoscopic behavior of superconductors \cite{ginzburg_theory_2009}. Its time-dependent extension, the TDGL theory, has been extensively applied to model transport properties and other nonequilibrium phenomena in superconducting systems \cite{hernandez_surface_2002, berdiyorov_finite-size_2009, bespalov_mismatch_2012, chung_numerical_2012, berdiyorov_dynamic_2012, koshelev_optimization_2016, sugai_driving_2021, motta_metamorphosis_2021, cornfeld_spin-polarized_2021, kapustin_universal_2023, cadorim_harnessing_2024, xue_case_2024}. Numerous studies have confirmed that TDGL simulations are in excellent agreement with experimental observations \cite{vodolazov_masking_2005, carty_visualizing_2008, berdiyorov_large_2012, pack_vortex_2020, lyu_superconducting_2021, he_revealing_2024}.
	
	The schematic setup in Fig.~\ref{fig_model} illustrates an asymmetric multilayer heterostructure, where the three constituent materials—Nb, V, and Ta—are periodically stacked along the $z$-direction with equal thicknesses, similar to the structure designed by Ando \textit{et al.}~\cite{ando_observation_2020}, but with thicker layers. In this extended multilayer system, microscopic quantum effects are effectively averaged out, allowing the macroscopic superconducting behavior to be captured within the TDGL framework. Furthermore, such multilayer superconducting structures may improve experimental feasibility. This model thus offers both theoretical consistency and practical relevance.
	
	Our results show that vortex dynamics, driven by the interplay between Lorentz forces and material-dependent preferences, plays a key role in the SDE in multilayer heterostructures under an external magnetic field.

	\section{Results}
	
	\subsection{Theoretical Model} 
	
	\begin{table}[tbp] 
		\caption{Superconducting parameters.}\label{tab1} 
		\begin{tabular}{@{}ccccccc@{}} 
			\toprule 
			& & $T_\mathrm{c}$ (K) & $\kappa$ & $\xi(0)$ (nm) & $\sigma$ ($10^6$ $S m^{-1}$) & Critical field (mT) \\ 
			\midrule
			& Ta & 4.5 & 0.30 & 95 & 7.4 & 83.5 ($H_\mathrm{c}$) \\ 
			& V  & 5.4 & 0.85 & 44 & 5.1 & 296.0 ($H_\mathrm{c2}$) \\ 
			& Nb & 9.2 & 0.90 & 39 & 6.6 & 390.0 ($H_\mathrm{c2}$) \\ 
			\botrule 
		\end{tabular} \\ 
		\textbf{Source:} Refs.~\cite{sekula_magnetic_1972, poole_handbook_2000, takata_magnetic_2015, hu_superconducting_2018} 
	\end{table}
		
	The conventional superconductors considered here have relatively low Ginzburg-Landau parameters $\kappa$, defined as the ratio of the penetration depth $\lambda$ to the coherence length $\xi$, as shown in Table \ref{tab1}. A small value of $\kappa$ implies a large coherence length and a high condensation energy, which suppresses thermal fluctuations and reduces the relevance of tilt deformations, in contrast to strongly type-II superconductors. Thus, rigid vortex lines aligned with the direction of an external magnetic field allow the vortex system to be treated as effectively two-dimensional. This approximation is analogous to studies of three-dimensional vortices aligned with correlated columnar defects, which can be reduced to an effective two-dimensional description in the plane perpendicular to the defects \cite{blatter_vortices_1994}. The sample thickness is $120 \,\mathrm{nm}$ along the $z$-direction, as reported by Ando \textit{et al.}~\cite{ando_observation_2020}, while its length along the magnetic-field direction ($y$-axis) extends to several hundred micrometers. Thus, the sample can be reasonably approximated as infinitely long along the magnetic field direction. Consequently, the simulated region is small relative to the entire sample and can be effectively reduced to a two-dimensional model \cite{wang_effects_2022}, specifically the $xz$ cross-section. On the basis of these two considerations, all TDGL simulations in this work are carried out in two dimensions for simplicity. The system is modeled as a strip with dimensions $L_x$ (length) and $L_z$ (width). An external current $\mathbf{j_\mathrm{ext}}$ is applied along the $x$-direction, while an external magnetic field $\mathbf{H_\mathrm{ext}}$ is applied perpendicular to the plane, along the $y$-axis. The positive current direction is defined as $+x$. For clarity, a positive (negative) external current along $+x$ ($-x$) is denoted as $+j$ ($-j$) throughout this paper.
	
	In the TDGL framework, a superconductor is described by a complex order parameter $\psi$, where $\lvert \psi \rvert^{2}$ denotes the local Cooper pair density (CPD). For convenience, the subscripts $1$, $2$, and $3$ are used to label Nb, V, and Ta, respectively. The TDGL equations, expressed in the zero-electric-potential gauge, are given by:
	\begin{eqnarray}
		\frac{\hbar^2}{2 m_\star D_i} \frac{\partial \psi}{\partial t} &=& \left[ \frac{\hbar^2}{2 m_\star} \left( \nabla - i \frac{e_\star}{\hbar} \mathbf{A} \right)^2 + \alpha_i - \beta_i |\psi|^2 \right] \psi, \nonumber \\
		\sigma_i \frac{\partial \mathbf{A}}{\partial t} &=& \frac{\hbar e_\star}{2 i m_\star} \left( \psi^{\star} \nabla \psi - \psi \nabla \psi^\star \right) - \frac{e_\star^2}{m_\star} |\psi|^2 \mathbf{A} \nonumber \\
		&& - \frac{1}{\mu_0} \nabla \times \left( \nabla \times \mathbf{A} - \mathbf{H}_\mathrm{ext} \right), \label{eq_A0}
	\end{eqnarray}
	where the subscript $i \in \{1,2,3\}$ specifies the corresponding layer. $D_i$ and $\sigma_i$ denote the phenomenological normal-state diffusion constant and electrical conductivity, respectively. These parameters can be derived from microscopic theory and are related by $D_i \propto \kappa_i^{-2} \sigma_i^{-1}$ \cite{kato_effects_1993, hernandez_surface_2002}. The magnetic field is obtained from the curl of the vector potential, $\nabla \times \mathbf{A} = \mathbf{H}$. The quantities $e_\star$ and $m_\star$ represent the effective charge and mass of Cooper pairs, respectively. Equation~\eqref{eq_A0} corresponds to the Maxwell equation governing the magnetic field and total current. The phenomenological parameters $\alpha_i$ and $\beta_i$ are defined as
	\begin{eqnarray}
		\alpha_i = \frac{\hbar^2}{2 m_\star \xi_i^2}, \quad 
		\beta_i = \mu_0 \frac{\hbar^2 e_\star^2 \kappa_i^2}{2 m_\star^2}.
	\end{eqnarray}
	Both $\alpha_i$ and the coherence length $\xi_i$ depend on temperature $T$ according to
	\begin{eqnarray}
		\alpha_i (T) = \alpha_i (0) \left( 1 - \frac{T}{T_{\mathrm{c},i}} \right), \quad
		\xi_i (T) = \xi_i(0) \left( 1 - \frac{T}{T_{\mathrm{c},i}} \right)^{-\frac{1}{2}},
	\end{eqnarray}
	where $T_{\mathrm{c},i}$ denotes the critical temperature of each material.
	
	It is well established that the TDGL equations can be reformulated in a dimensionless form. Because the layer thickness in our simulation is much larger than the $1.0~\mathrm{nm}$ used by Ando \textit{et al.}~\cite{ando_observation_2020}, we do not adopt the coherence length ($13~\mathrm{nm}$) associated with the $\mathrm{[Nb(1.0\,nm)/V(1.0\,nm)/Ta(1.0\,nm)]}_{40}$ superlattice. Instead, we use the respective bulk superconducting parameters, summarized in Table~\ref{tab1}. For the multimaterial system considered here, we reformulate the TDGL equations in dimensionless form following the approach of Wang \textit{et al.} \cite{wang_effects_2022}.
		
	The dimensionless units are defined with respect to Nb, which has the highest critical temperature and the largest upper critical field among the three materials. The length scale is set by the zero-temperature coherence length, $\xi_0 = \xi_{1}(0)$, while the time scale is given by the relaxation time of the GL order parameter, $\tau = \xi_0^2 / D$. Temperature is normalized by the critical temperature, $T_\mathrm{c}=T_{\mathrm{c},1}$; the magnetic field by the upper critical field, $H_\mathrm{c2}= \kappa_1 \sqrt{2 \mu_0 \alpha_1^2 / \beta_1}$; the order parameter $\psi$ by $\psi_0 = \sqrt{\alpha_1 / \beta_1}$; and the conductivity by the normal-state value, $\sigma_0 = (\mu_0 D_1 \kappa_1^2)^{-1}$. Setting $\sigma_1 = \sigma_0$, the TDGL equations in dimensionless form are
	\begin{eqnarray} 
		&&\frac{\partial \psi}{\partial t} = \left[ \left( \nabla - i \mathbf{A} \right)^2 + 1 - T - |\psi|^2 \right] \psi, \nonumber \\ 
		&&\frac{\partial \mathbf{A}}{\partial t} = \operatorname{Im} \left[ \psi^{\star} (\nabla - i \mathbf{A}) \psi \right] - \kappa_1^{2} \nabla \times \left( \nabla \times \mathbf{A} - \mathbf{H}_\mathrm{ext} \right), \nonumber \\
		\label{eq_Nb} 
	\end{eqnarray} 
	for the Nb layer, which corresponds to the standard TDGL equations for isotropic superconductors, and
	\begin{eqnarray} 
		&&\frac{\sigma_i}{\sigma_1} \frac{\partial \psi}{\partial t} = \left[ \frac{\kappa_1^2}{\kappa_i^2} \left( \nabla - i \mathbf{A} \right)^2 + \frac{\kappa_1^2 \xi_0^2}{\kappa_i^2 \xi_i^2} \left( 1 - T \frac{T_\mathrm{c}}{T_{\mathrm{c},i}} \right) - |\psi|^2 \right] \psi, \nonumber \\ 
		&&\frac{\sigma_i}{\sigma_1} \frac{\partial \mathbf{A}}{\partial t} = \operatorname{Im} \left[ \psi^{\star} (\nabla - i \mathbf{A}) \psi \right] - \kappa_1^{2} \nabla \times \left( \nabla \times \mathbf{A} - \mathbf{H}_\mathrm{ext} \right), \nonumber \\
			\label{eq_Ta_V} 
	\end{eqnarray} 
	for the Ta and V layers($i = \{ 2,3 \}$). In this study, we focus on superconducting properties relevant to the SDE mechanism, assuming that only $\kappa_i$, $\xi_i$, $\sigma_i$, and $T_{\mathrm{c},i}$ vary among materials. Other properties, such as thermal conductivity, are treated as uniform. Notably, substituting $D_i \propto \kappa_i^{-2} \sigma_i^{-1}$ shows that the TDGL equations for the asymmetric heterostructure presented here are equivalent to those given by Wang \textit{et al.} \cite{wang_effects_2022}.
	
	Since the SDE typically emerges near the critical current, vortex motion can become highly dynamic, leading to substantial Joule heating. To capture this effect, the present TDGL simulations are coupled with a heat-transfer equation, expressed as
	\begin{equation} 
		\nu \, \partial_t T = \zeta \nabla^2 T + \sigma_i \left( \frac{\partial \mathbf{A}}{\partial t} \right)^2 + \eta (T_0 - T),
		\label{eq_T} 
	\end{equation}
	where $T_0$ denotes the bath temperature, $\nu = 0.03$ is the heat capacity of the sample, and $\zeta = 0.06$ represents its thermal conductivity. The parameter $\eta = 2 \times 10^{-4}$ quantifies the efficiency of heat exchange between the sample and its holder. These phenomenological coefficients correspond to moderate heat dissipation conditions, which have been validated as appropriate for modeling typical experimental environments \cite{vodolazov_masking_2005, berdiyorov_large_2012, lyu_superconducting_2021, cadorim_harnessing_2024}.
	
	The iterative Jacobi method \cite{sadovskyy_stable_2015} is employed to solve Eqs.~\eqref{eq_Nb}, \eqref{eq_Ta_V}, and \eqref{eq_T}. Periodic boundary conditions are imposed along the $x$-direction, aligned with the external current, while Neumann boundary conditions are applied along the transverse $z$-direction. The external current is introduced through a current-induced magnetic field $H_I$, such that the boundary conditions satisfy $H\big|_{z=0} = H_\mathrm{ext} + H_I$ and $H\big|_{z=L_z} = H_\mathrm{ext} - H_I$. In this study, $H_I$ is gradually increased in small increments of $\Delta H_I = 0.01$, and the system is evolved for $10^4 \tau$ at each discrete value of $H_I$ to ensure that it reaches a steady state. The simulations are implemented in C++ to enable high-performance parallel computation at each discretized grid point, using the Compute Unified Device Architecture (CUDA) on an NVIDIA RTX 3090 GPU.
	
	\subsection{Numerical Simulation} 
	
	\subsubsection{SDE in a 3-cycle periodic layer stack} 
	
	\begin{figure}[tbp]
		\includegraphics[width=1.0\linewidth]{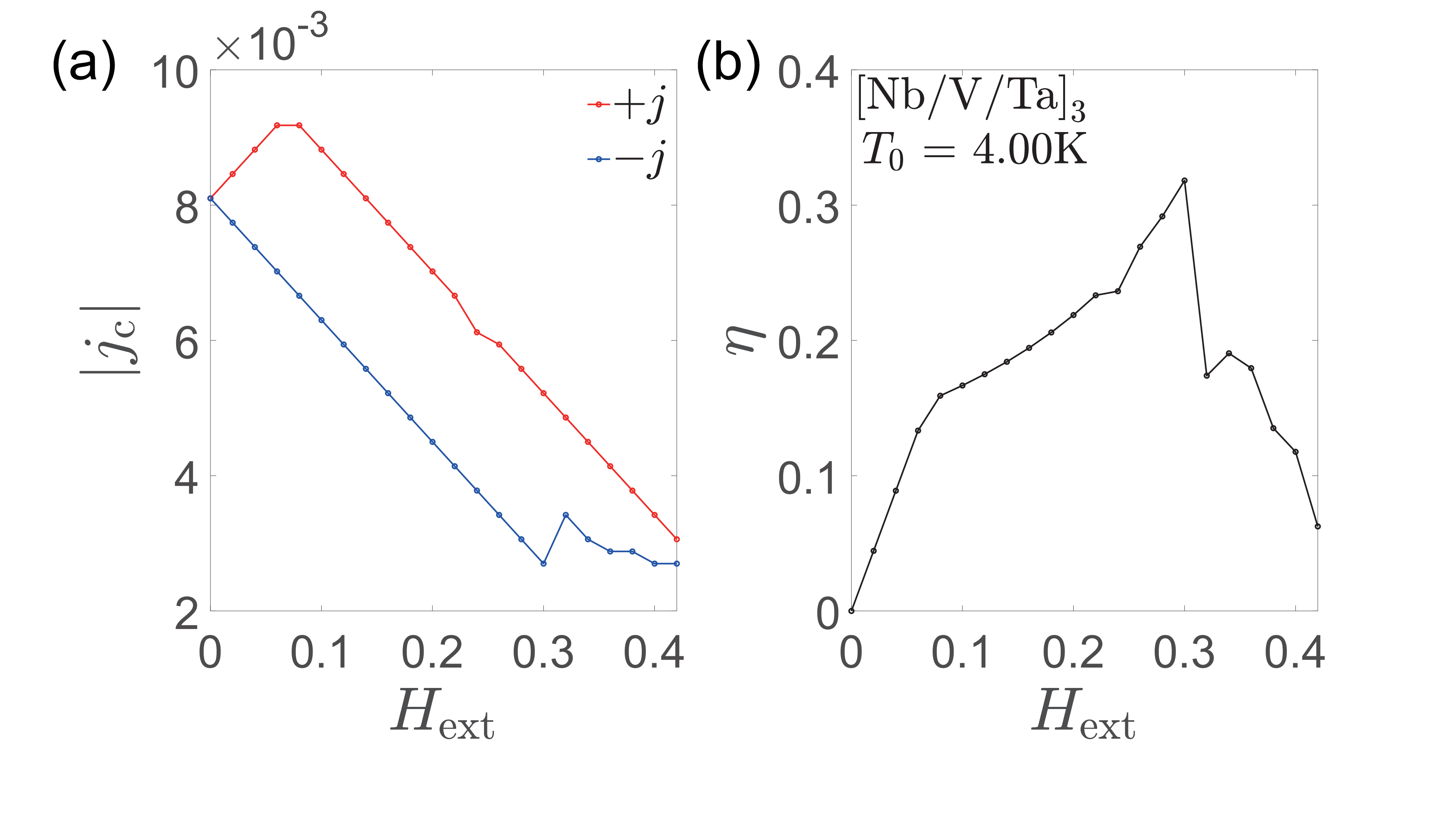}
		\caption{
			\textbf{Critical current and superconducting diode efficiency in a 3-cycle periodic layer stack.}
			\textbf{(a)} Critical current $\lvert j_\mathrm{c}^\pm \rvert$ as a function of the external magnetic field. The red line and blue line correspond to the positive and negative current directions ($\pm j$), respectively, and the vertical axis shows the magnitude of the critical current in each case. 
			\textbf{(b)} Superconducting diode efficiency $\eta$ as a function of the external magnetic field. The black line indicates the efficiency. All curves are obtained for the multilayer sample at a bath temperature of $T_0 = 4.00$ K.
		}
		\label{Jc_3layers}
	\end{figure}
	
	\begin{figure*}[tbp]
		\centering
		\includegraphics[width=1.0\linewidth]{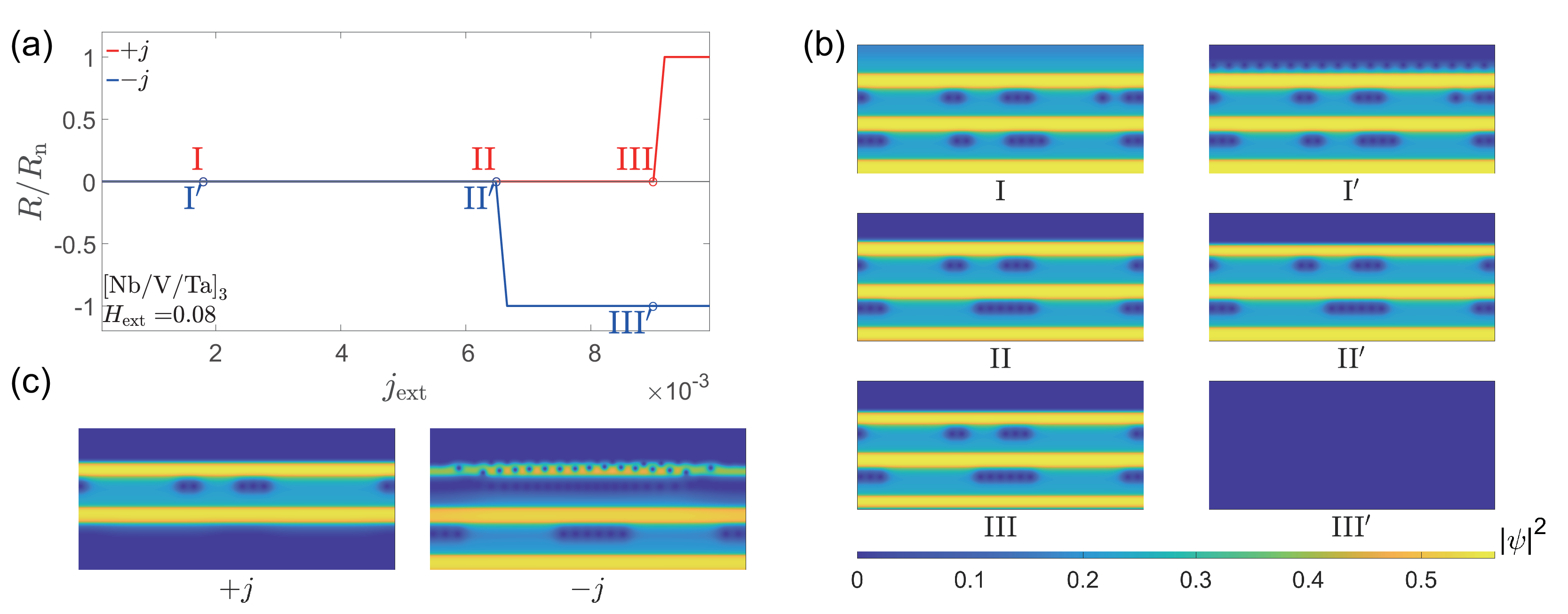}
		\caption{
			\textbf{Current–resistance behavior and Cooper-pair density distributions in a 3-cycle periodic layer stack.}
			\textbf{(a)} Current–resistance characteristics of the sample at a bath temperature of $T_0 = 4.0~\mathrm{K}$ and an external magnetic field of $H_\mathrm{ext} = 0.08H_\mathrm{c2}$. The resistance is normalized by the normal-state resistance $R_\mathrm{n}$. The red line corresponds to the positive current direction ($+j$) and the blue line corresponds to the negative current direction ($-j$). Points labelled I, II, and III on the red curve and I$'$ , II$'$ , and III$'$ on the blue curve mark representative states used in the (c), where the applied currents have equal magnitude but opposite directions.
			\textbf{(b)} Distributions of the Cooper-pair density for the states indicated in (a). Colors represent the magnitude of the density, with blue indicating low values and yellow indicating high values along the color bar. 
			\textbf{(c)} Cooper-pair density distributions during the phase transition for positive and negative current directions. 
			The full phase-transition process is provided in Supplementary Movie~1.
		}
		\label{fig_3layers}
	\end{figure*}
	
	\begin{figure}[tbp]
		\includegraphics[width=1.0\linewidth]{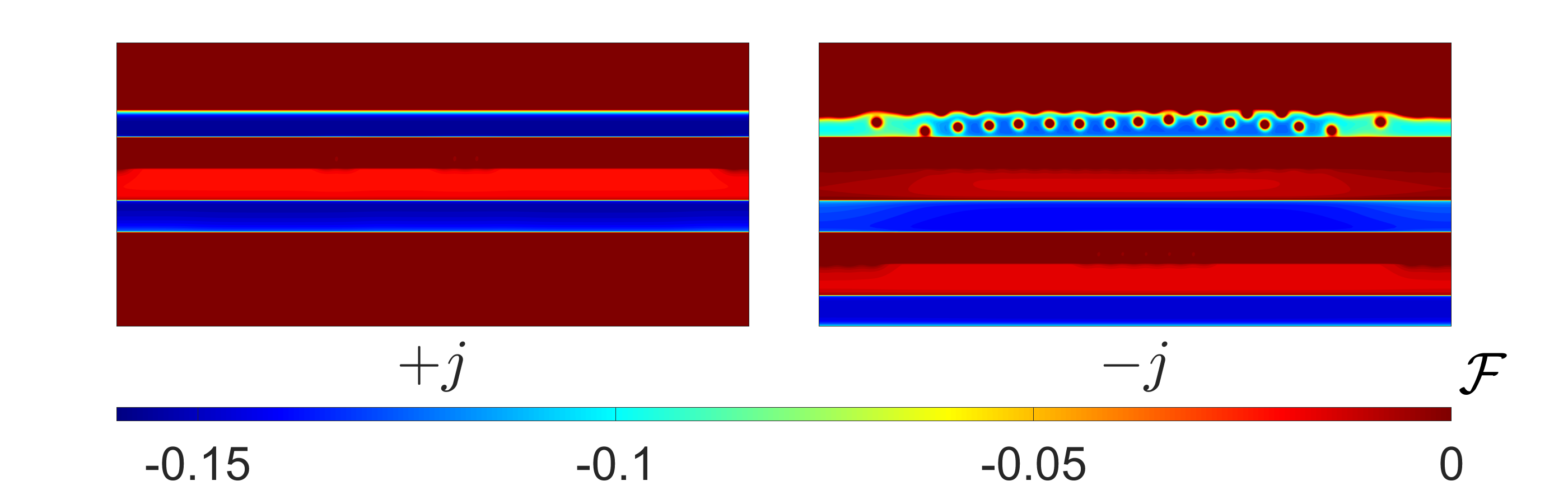}
		\caption{
			\textbf{Free-energy distributions during the phase transition.}
			Free-energy $\mathcal{F}$ distributions during the phase transition for the positive and negative current directions ($\pm j$). Colors indicate the magnitude of the free energy, with blue representing low values and red representing high values along the color bar. The two distributions correspond to the states shown in (c) of Fig.~3.}
		\label{free_3layers}
	\end{figure}
	
	\begin{figure*}[tbp]
		\includegraphics[width=1.0\linewidth]{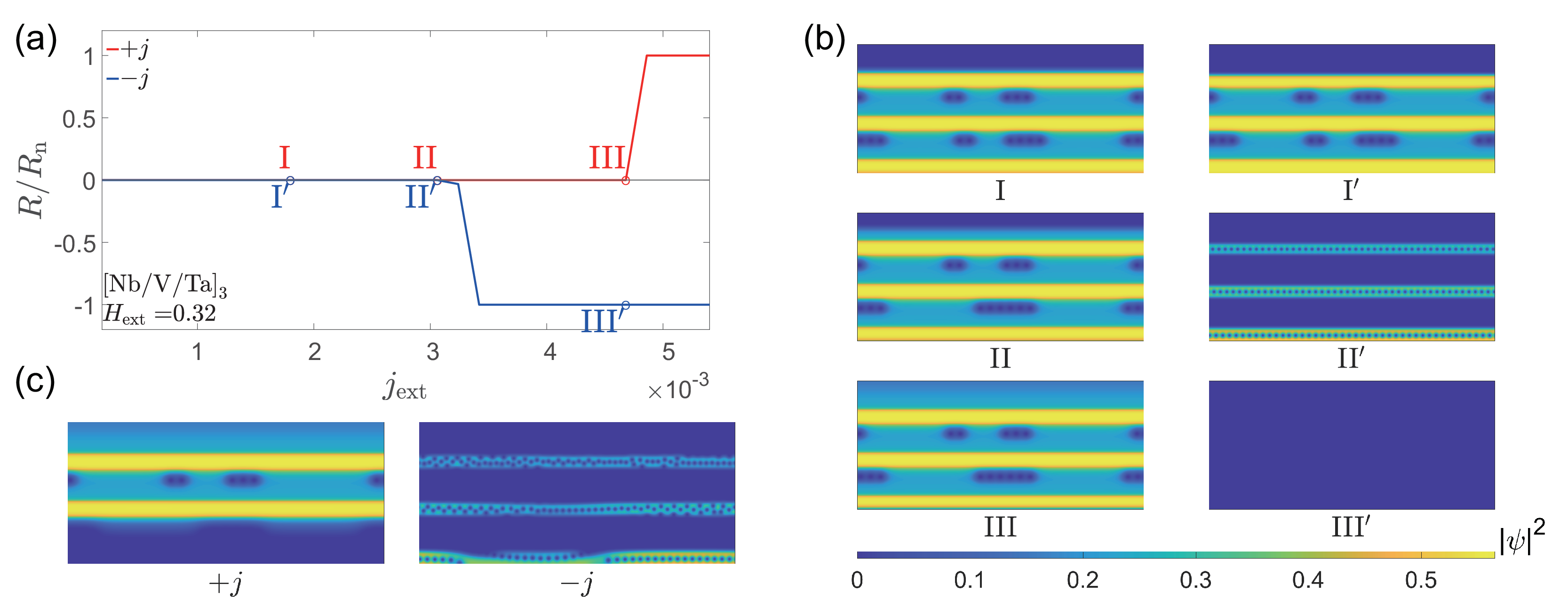}
		\caption{
			\textbf{Current–resistance behavior and Cooper-pair density distributions at higher magnetic field.}
			\textbf{(a)} Current–resistance characteristics of the sample at a bath temperature of $T_0 = 4.0~\mathrm{K}$ and an external magnetic field of $H_\mathrm{ext} = 0.32H_\mathrm{c2}$. The resistance is normalized by the normal-state resistance $R_\mathrm{n}$. The red line corresponds to the positive current direction ($+j$) and the blue line corresponds to the negative current ($-j$) direction. Points labelled I, II, and III on the red line and I$'$, II$'$, and III$'$ on the blue line indicate representative states used in the (c), where the applied currents have equal magnitude but opposite sign. 
			\textbf{(b)} Distributions of the Cooper-pair density for the states marked in (a). Colors represent the magnitude of the density, with blue indicating low values and yellow indicating high values along the color bar. 
			\textbf{(c)} Cooper-pair density distributions during the phase transition for the positive and negative current directions. 
			The complete phase-transition process is shown in Supplementary Movie~2.
		}
		\label{fig_3layers_-}
	\end{figure*}
	
	We first configure the multilayer sample as $\mathrm{[Nb(10\xi_0)/V(10\xi_0)/Ta(10\xi_0)]}_3$, where each layer has a thickness of $10\xi_0$. This cyclic stacking structure is repeated three periods, giving a total thickness of $L_z = 90\xi_0$. The sample length along the $x$-direction is set to $L_x = 200\xi_0$. The bath temperature is fixed at $T_0 = 4.0~\mathrm{K}$, which is below the critical temperatures of all three superconducting materials.

	We first examine the overall behavior of the critical currents $j_\mathrm{c}^\pm$ and the corresponding superconducting diode efficiency $\eta$,
		\begin{equation}
			\eta = \frac{j_\mathrm{c}^+ - \lvert j_\mathrm{c}^- \rvert}{j_\mathrm{c}^+ + \lvert j_\mathrm{c}^- \rvert},
		\end{equation}
	as functions of $H_\mathrm{ext}$. As shown in Fig. \ref{Jc_3layers} , a matching field for $+j$ appears near $H_\mathrm{ext} = 0.08 H_\mathrm{c2}$, where $j_\mathrm{c}^+$ first reaches a peak and then decreases almost linearly with further increases in $H_\mathrm{ext}$. For $-j$, a matching field also occurs near $H_\mathrm{ext} = 0.32 H_\mathrm{c2}$; however, because this field is relatively large, it produces only a local maximum of $\lvert j_\mathrm{c}^- \rvert$. In most regions before and after this field, $\lvert j_\mathrm{c}^- \rvert$ decreases with increasing $H_\mathrm{ext}$, although the decay rate above the matching field is smaller than that below it. Over the range $0.08 H_\mathrm{c2} \leqslant H_\mathrm{ext} \leqslant 0.30 H_\mathrm{c2}$, $\lvert j_\mathrm{c}^\pm \rvert$ decrease at nearly the same rate, resulting in a maximum efficiency of $\eta \simeq 31.8\%$ just before the matching field of $-j$.
	
	Accordingly, we examine the current–resistance characteristics of the system near the matching field for $+j$ at $H_\mathrm{ext} = 0.08 H_\mathrm{c2}$, as shown in Fig.~\ref{fig_3layers}(a). For $-j$, the transition to the normal state occurs at $j_\mathrm{c}^- \simeq -0.0065$, whereas for $+j$, the system remains superconducting up to $j_\mathrm{c}^+ \simeq 0.0090$. The resulting nonreciprocal critical current density is $\Delta j_\mathrm{c} = j_\mathrm{c}^+ - \lvert j_\mathrm{c}^- \rvert \simeq 0.0025$, yielding $\eta \simeq 16.1\%$. Notably, the flux-flow regime is nearly absent for both current directions, indicating a direct transition from the superconducting to the normal state at their respective critical currents, without an intermediate dissipative phase. Consequently, the superconducting diode exhibits minimal energy loss and excellent operational performance.

	The distribution of the CPD provides important insights into the mechanism underlying the SDE. As shown in Fig.~\ref{fig_3layers}(b), vortices appear exclusively within the Ta layer at all points except for point I$'$. At point I$'$, vortices are also observed in the V layer; however, their sizes vary depending on the coherence length $\xi$. Under zero-field conditions, the CPD in each layer can be estimated as $\frac{\kappa_1^2 \xi_0^2}{\kappa_i^2 \xi_i^2} \left(1 - \frac{T}{T_{\mathrm{c},i}} \right)$: 0.57 for Nb, 0.23 for V, and 0.17 for Ta. These values are consistent with the CPD observed under a weak magnetic field (point I). A common feature in both cases is that the Ta layer at the upper boundary transitions to the normal state earlier, owing to its lowest critical field and its type-I superconducting nature ($\kappa < 0.707$). Meanwhile, vortices remain confined within individual layers even immediately before the phase transition (see points III and II$'$), effectively acting as quasi-independent subsystems.
	
	As shown in Fig.~\ref{fig_3layers}(c) and Supplementary Movie 1, numerous vortices emerge during the phase transition under $-j$. Their vigorous motion generates substantial Joule heating, which raises the sample temperature and drives it into the normal state. In contrast, during the phase transition under $+j$, vortex motion is scarcely observed. Consequently, $\lvert j_\mathrm{c}^- \rvert$ is markedly smaller than $\lvert j_\mathrm{c}^+ \rvert$. Because the CPD vanishes at the vortex core, a larger vortex can be regarded as carrying more energy. Since vortices tend to move in order to minimize the extent of their cores, and given that the coherence lengths are 95~nm for Ta, 44~nm for V, and 39~nm for Nb, vortices are expected to propagate sequentially from Ta to V and then to Nb.
	
	Furthermore, since the TDGL framework is derived from a free-energy functional, analyzing the distribution of free energy
		\begin{eqnarray}
			\mathcal{F} = &&- \alpha \lvert \psi \rvert^2 + \frac{\beta}{2} \left \vert \psi \right \vert^4 + \frac{\hbar^2}{2 m_\star} \left \vert \left( \nabla - i \frac{e_\star}{\hbar} \mathbf{A} \right) \psi \right \vert^2 \nonumber \\
			&& + \frac{1}{2 \mu_0} \lvert \nabla \times \mathbf{A} - \mathbf{H}_\mathrm{ext} \rvert^2
		\end{eqnarray}
	provides important insights into the observed nonreciprocity. Although the coefficients in $\mathcal{F}$ differ among the layers, the free energy of Nb remains significantly lower than that of the other two materials because its CPD is much larger, particularly when the Ta and V layers are in the normal state, as shown in Fig.~\ref{free_3layers}. The free energy of V is slightly lower than that of Ta. Even under zero field, the estimated CPD values already indicate that the free energy of Nb is much lower than that of Ta and V, with CPD values of 0.57 for Nb, 0.23 for V, and 0.17 for Ta.
	
	The essence of vortex motion is to reduce the system’s total free energy $\mathcal{F}$ through relaxation; thus, vortices spontaneously move from high-energy to low-energy regions, namely from Ta to V and then to Nb. When $+j$ is applied, the Lorentz force on the vortices points along the $+z$ direction, opposite to this tendency of vortices motion, thereby suppressing vortex dynamics. As a result, no vortex-flow state develops, and the superconducting–normal transition occurs directly. In contrast, under $-j$, the Lorentz force is oriented along the $-z$ direction, consistent with this tendency of vortices motion. This enhances vortex activity and significantly reduces the critical current. The intense vortex motion also induces rapid heating of the sample, rendering the vortex-flow state extremely short-lived.

	For completeness, we also examined the behavior near the matching field $H_\mathrm{ext} = 0.32H_\mathrm{c2}$ under $-j$. As shown in Fig.~\ref{fig_3layers_-}(a), $j_\mathrm{c}^+ \simeq 0.0047$ while $\lvert j_\mathrm{c}^- \rvert \simeq 0.0032$, yielding $\eta \simeq 19.0\%$. In contrast to Fig.~\ref{fig_3layers}(b), a large number of vortices are already present in the system before the phase transition under $-j$ (point II$'$). These vortices are stably confined within the Nb layer and are considerably smaller than those observed in the Ta layer, as also clearly shown in the Supplementary Movie 2.
	
	This behavior can be understood as follows: before reaching the matching field, the current required to nucleate vortices is relatively large. Once vortices begin to move, substantial Joule heating rapidly raises the sample temperature, driving it into the normal state. Beyond the matching field, however, the current needed to nucleate vortices is much smaller, leading to a weaker Lorentz force. This allows a stable vortex state to form, thereby enhancing $\lvert j_\mathrm{c}^- \rvert$. Since the Lorentz force under $-j$ still aligns with the preferred tendency of vortices motion, the SDE persists with the same polarity.
	
	\subsubsection{SDE in a 15-cycle periodic layer stack}

	\begin{figure}[tbp]
		\includegraphics[width=1.0\linewidth]{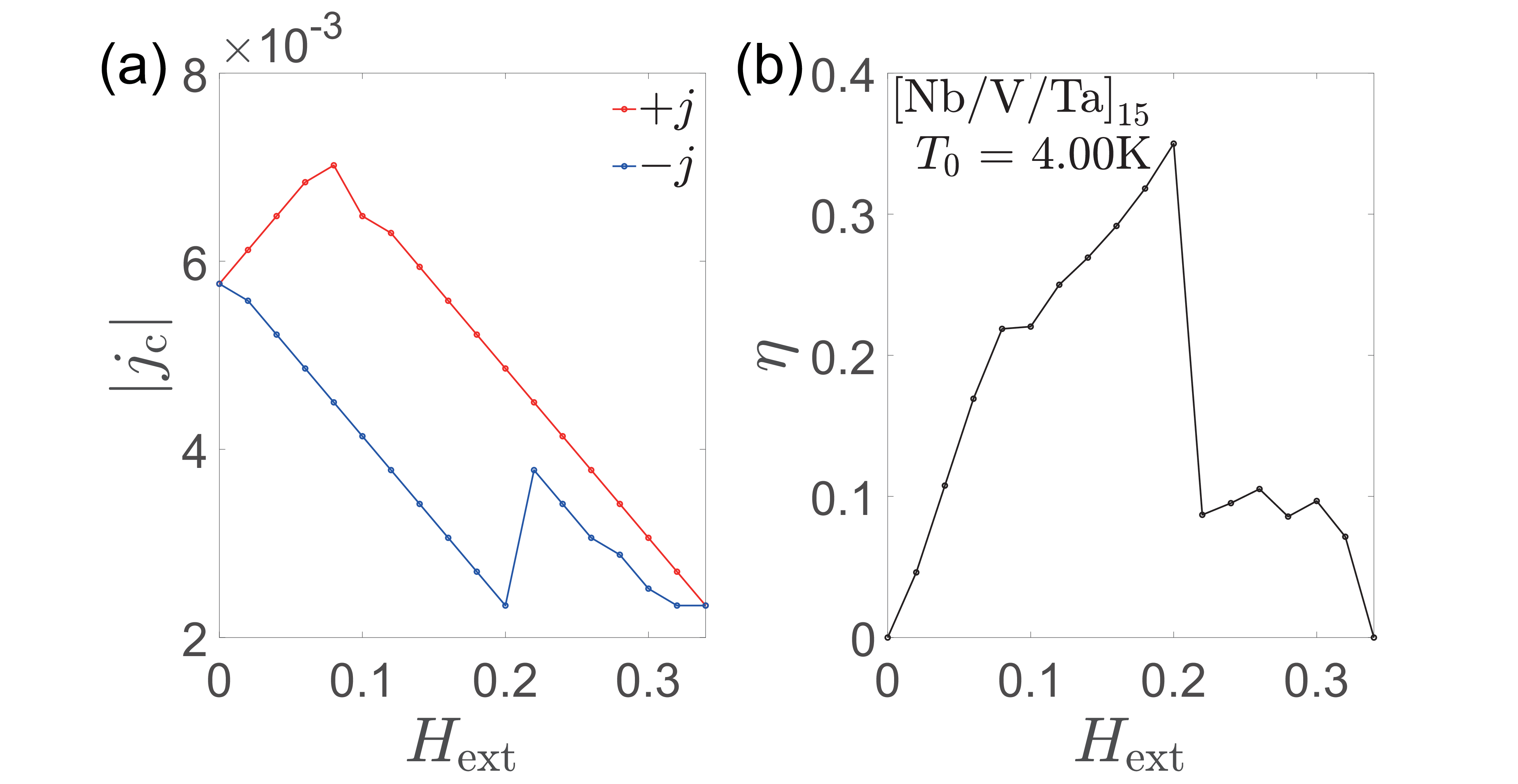}
		\caption{
			\textbf{Critical current and superconducting diode efficiency in a 15-cycle periodic layer stack.}
			\textbf{(a)} Critical current $\lvert j_\mathrm{c}^\pm \rvert$ as a function of the external magnetic field for the multilayer sample at a bath temperature of $T_0 = 4.0~\mathrm{K}$. The red line corresponds to the positive current direction ($+j$) and the blue line corresponds to the negative current direction ($-j$). 
			\textbf{(b)} Superconducting diode efficiency $\eta$ as a function of the external magnetic field. The black line indicates the efficiency. 
		}
		\label{Jc_15layers}
	\end{figure}
	
	\begin{figure*}
		\centering
		\includegraphics[width=1.0\linewidth]{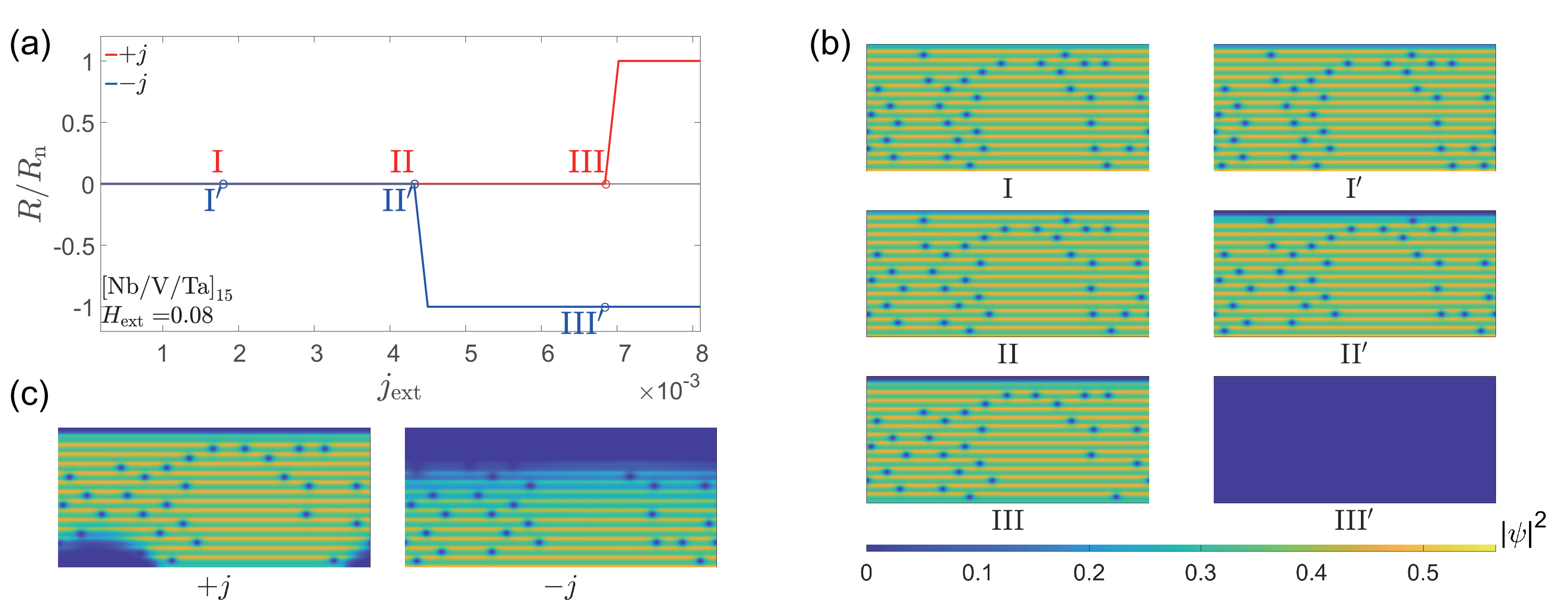}
		\caption{
			\textbf{Current–resistance behavior and Cooper-pair density distributions in a 15-cycle periodic layer stack.}
			\textbf{(a)} Current–resistance characteristics of the multilayer sample at a bath temperature of $T_0 = 4.0~\mathrm{K}$ and an external magnetic field of $H_\mathrm{ext} = 0.08H_\mathrm{c2}$. The resistance is normalized by the normal-state resistance $R_\mathrm{n}$.
			The red line corresponds to the positive current direction ($+j$) and the blue line corresponds to the negative current direction ($-j$). Points labelled I, II, and III on the red line and I$'$, II$'$, and III$'$ on the blue line indicate representative states used in the (c), where the applied currents have equal magnitude but opposite sign. 
			\textbf{(b)} Distributions of the Cooper-pair density for the states marked in (a). Colors represent the magnitude of the density, with blue indicating low values and yellow indicating high values along the color bar. 
			\textbf{(c)} Cooper-pair density distributions during the phase transition for the positive and negative current directions.  
			The complete phase-transition process is shown in Supplementary Movie~3.
		}
		\label{fig_15layers}
	\end{figure*}
	
	\begin{figure}[tbp]
		\includegraphics[width=1.0\linewidth]{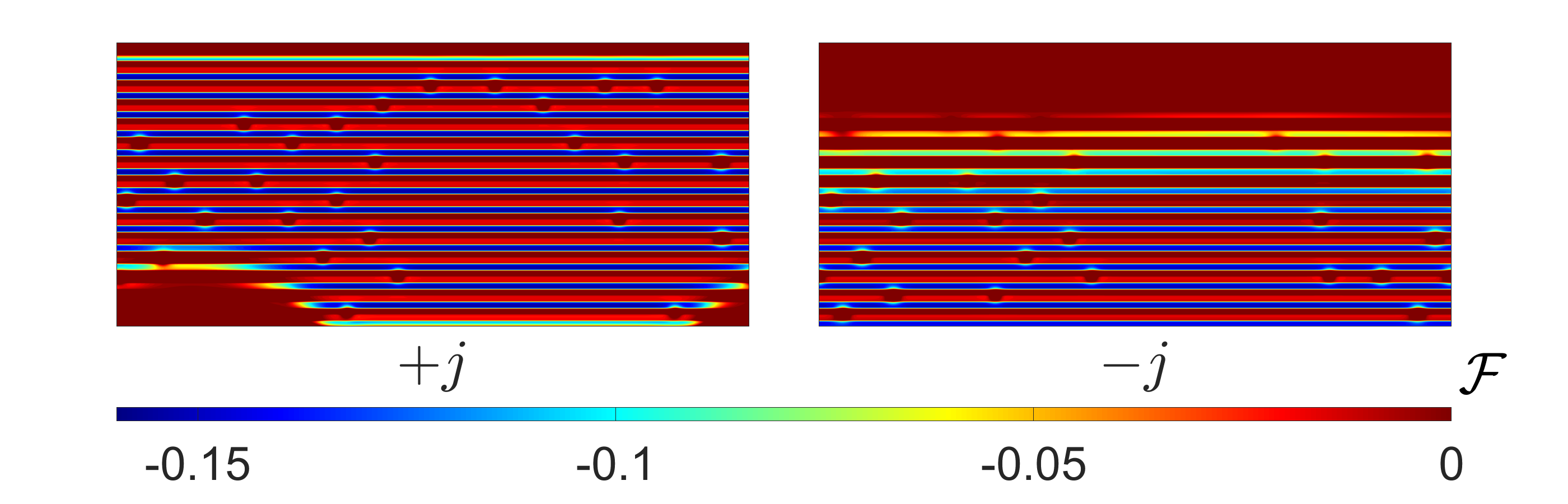}
		\caption{
			\textbf{Free-energy distributions during the phase transition.}
			Free-energy $\mathcal{F}$ distributions during the phase transition for the positive and negative current directions ($\pm j$). Colors indicate the magnitude of the free energy, with blue representing low values and red representing high values along the color bar. 
			These distributions correspond to the states shown in (c) of Fig.~7.}
		\label{free_15layers}
	\end{figure}
	
	\begin{figure*}[tbp]
		\centering
		\includegraphics[width=1.0\linewidth]{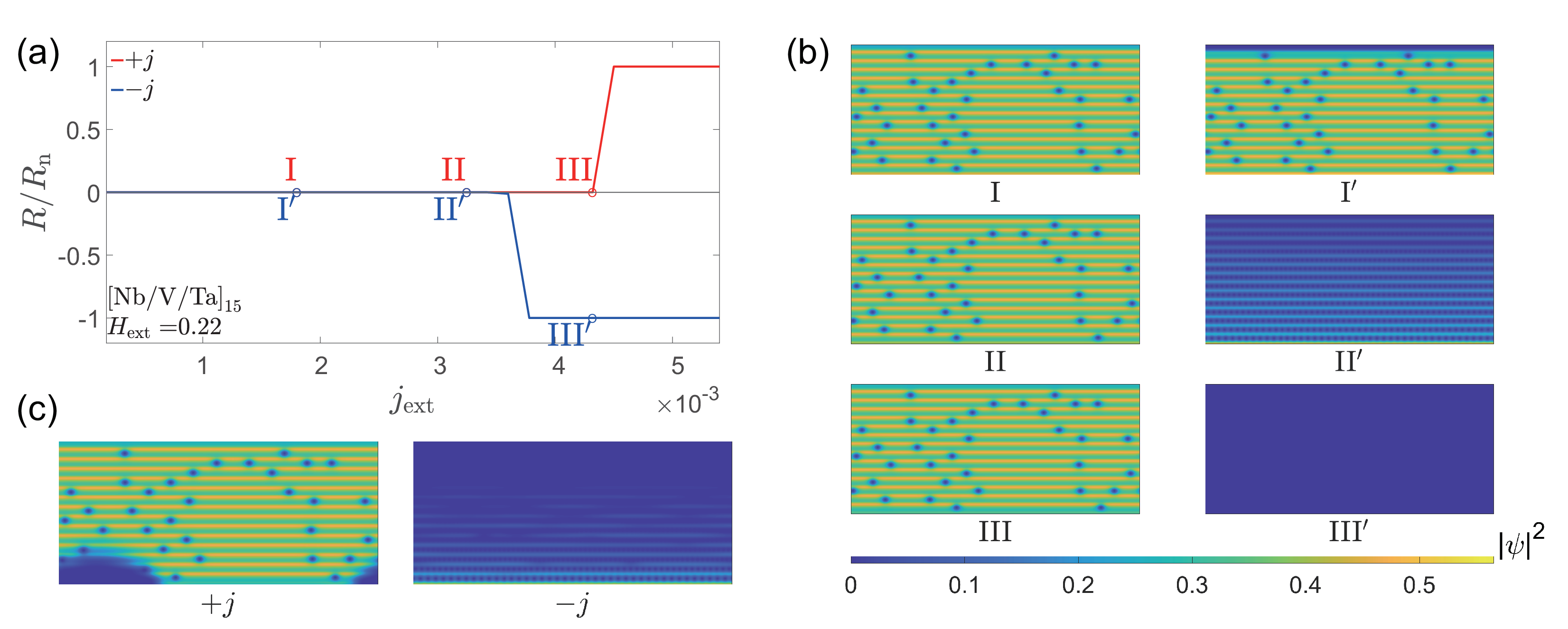}
		\caption{
			\textbf{Current–resistance behavior and Cooper-pair density distributions under higher magnetic field.}
			\textbf{(a)} Current–resistance characteristics of the multilayer sample at a bath temperature of $T_0 = 4.0~\mathrm{K}$ and an external magnetic field of $H_\mathrm{ext} = 0.22H_\mathrm{c2}$. The resistance is normalized by the normal-state resistance $R_\mathrm{n}$. The red line corresponds to the positive current direction ($+j$) and the blue line corresponds to the negative current direction ($-j$). Points labelled I, II, and III on the red line and I$'$, II$'$, and III$'$ on the blue line indicate representative states used in the (c), where the applied currents have equal magnitude but opposite sign. 
			\textbf{(b)} Distributions of the Cooper-pair density for the states marked in (a). Colors represent the magnitude of the density, with blue indicating low values and yellow indicating high values along the color bar. 
			\textbf{(c)} Cooper-pair density distributions during the phase transition for the positive and negative current directions. The complete phase-transition process is shown in Supplementary Movie 4.
		}
		\label{fig_15layers_-}
	\end{figure*}
	
	In this subsection, the layer thickness is reduced to $2\xi_0$, while the total thickness $L_z$, the length $L_x$, and the bath temperature remain unchanged. The sample structure is therefore defined as $\mathrm{[Nb(2\xi_0)/V(2\xi_0)/Ta(2\xi_0)]}_{15}$, corresponding to a cyclic multilayer sequence repeated over 15 periods. This configuration follows the approach of Ando \textit{et al.}~\cite{ando_observation_2020}, although the individual layer thicknesses in the present study are still considerably larger.
	
	Similarly, we examine the overall behavior of the critical currents $j_\mathrm{c}^\pm$ and the corresponding superconducting diode efficiency $\eta$ as functions of $H_\mathrm{ext}$. As shown in Fig.~\ref{Jc_15layers}, $j_\mathrm{c}^\pm$ and $\eta$ exhibit trends similar to those in $\mathrm{[Nb(10\xi_0)/V(10\xi_0)/Ta(10\xi_0)]}_3$, except that the absolute values of $\lvert j_\mathrm{c}^\pm \rvert$ are generally smaller. A matching field is observed near $H_\mathrm{ext} = 0.08 H_\mathrm{c2}$ for $+j$ and near $H_\mathrm{ext} = 0.22 H_\mathrm{c2}$ for $-j$. Moreover, the SDE disappears when $H_\mathrm{ext}$ reaches $0.34 H_\mathrm{c2}$, while the maximum efficiency of $\eta \simeq 35.0\%$ occurs just before the matching field of $-j$.
	
	Accordingly, we simulate the current–resistance characteristics of the system near the matching field for $+j$ at $H_\mathrm{ext} = 0.08H_\mathrm{c2}$, as shown in Fig.~\ref{fig_15layers}(a). For $-j$, the transition to the normal state occurs at $j_\mathrm{c}^- \simeq -0.0043$, whereas for $+j$, the superconducting state persists up to $j_\mathrm{c}^+ \simeq 0.0068$, yielding $\eta \simeq 22.5\%$. Notably, the flux-flow regime is nearly absent in both current directions, indicating a direct transition from the superconducting to the normal state at their respective critical currents.
	
	To further investigate the effects of reduced layer thickness, we analyze the CPD distributions. As shown in Fig.~\ref{fig_15layers}(b), thinner layers destabilize the ability of individual layers to accommodate a full row of vortices, thereby diminishing their quasi-independent behavior. Moreover, composite vortices spanning multiple layers begin to emerge (see points I and I$'$). During the phase-transition process [Fig.~\ref{fig_15layers}(c) and Supplementary Movie 3], a distinctly different behavior is observed: under $+j$, the normal state expands irregularly inward from the boundary, resembling the transition in an isotropic superconductor; by contrast, under $-j$, the normal state develops layer by layer, with vortex motion gradually penetrating the sample along the $-z$ direction.
	
	Since only composite vortices spanning multiple layers exist in this case, the difference in vortex core energy arising from variations in coherence length among the three materials is no longer critical. We therefore directly compute the distribution of the free energy $\mathcal{F}$. The distribution of $\mathcal{F}$ during the phase transition, shown in Fig.~\ref{free_15layers}, confirms that the $\mathcal{F}$ of Nb remains significantly lower than that of the other two materials, while the $\mathcal{F}$ of V is slightly lower than that of Ta. Consequently, the preferred tendency of vortices motion remains directed from Ta to V to Nb. Under $-j$, the Lorentz force is oriented along the $-z$ direction, consistent with this tendency of vortices motion. This alignment enhances vortex activity, leading to a pronounced reduction in $\lvert j_\mathrm{c}^- \rvert$ relative to $\lvert j_\mathrm{c}^+ \rvert$.
	
	For completeness, we also examined the behavior near the matching field $H_\mathrm{ext} = 0.22H_\mathrm{c2}$ under $-j$. As shown in Fig.~\ref{fig_15layers_-}(a), $j_\mathrm{c}^+ \simeq 0.0043$ while $\lvert j_\mathrm{c}^- \rvert \simeq 0.0036$, yielding $\eta \simeq 8.9\%$. In contrast to Fig.~\ref{fig_15layers}(b), a large number of vortices are already present in the system before the phase transition under $-j$ (point II$'$). Although the SDE persists with the same polarity, a key distinction from all previous CPD distributions emerges. At point II$'$ in Fig.~\ref{fig_15layers_-}, vortices are distributed across the entire sample under $-j$. Owing to the reduced layer thickness, these vortices readily form and traverse multiple layers, producing an effect reminiscent of a superlattice. As shown in the phase-transition process [Supplementary Movie 4 and Fig.~\ref{fig_15layers_-}(c)], vigorous vortex motion generates substantial Joule heating, rapidly driving the entire sample into the normal state and resulting in the near absence of a flux-flow regime. Under $+j$, by contrast, the normal state expands irregularly inward from the boundary, resembling the transition behavior of an isotropic superconductor.
	
	We further propose that as the layer thickness decreases, vortices spanning more layers form, while the preferred direction of vortex motion remains unchanged. Since the SDE observed in both asymmetric heterostructures over a wide magnetic-field range is consistent with that reported for the superlattice by Ando \textit{et al.}~\cite{ando_observation_2020}, it is plausible that the SDE in the $\mathrm{[Ta(1.0\,nm)/V(1.0\,nm)/Nb(1.0\,nm)]}_{40}$ superlattice is governed primarily by the same mechanism identified here. At the very least, the contribution of vortex dynamics to the SDE cannot be neglected.
	
	\subsubsection{Different stacking order} 
	
	\begin{figure*}[tbp]
		\centering
		\includegraphics[width=1.0\linewidth]{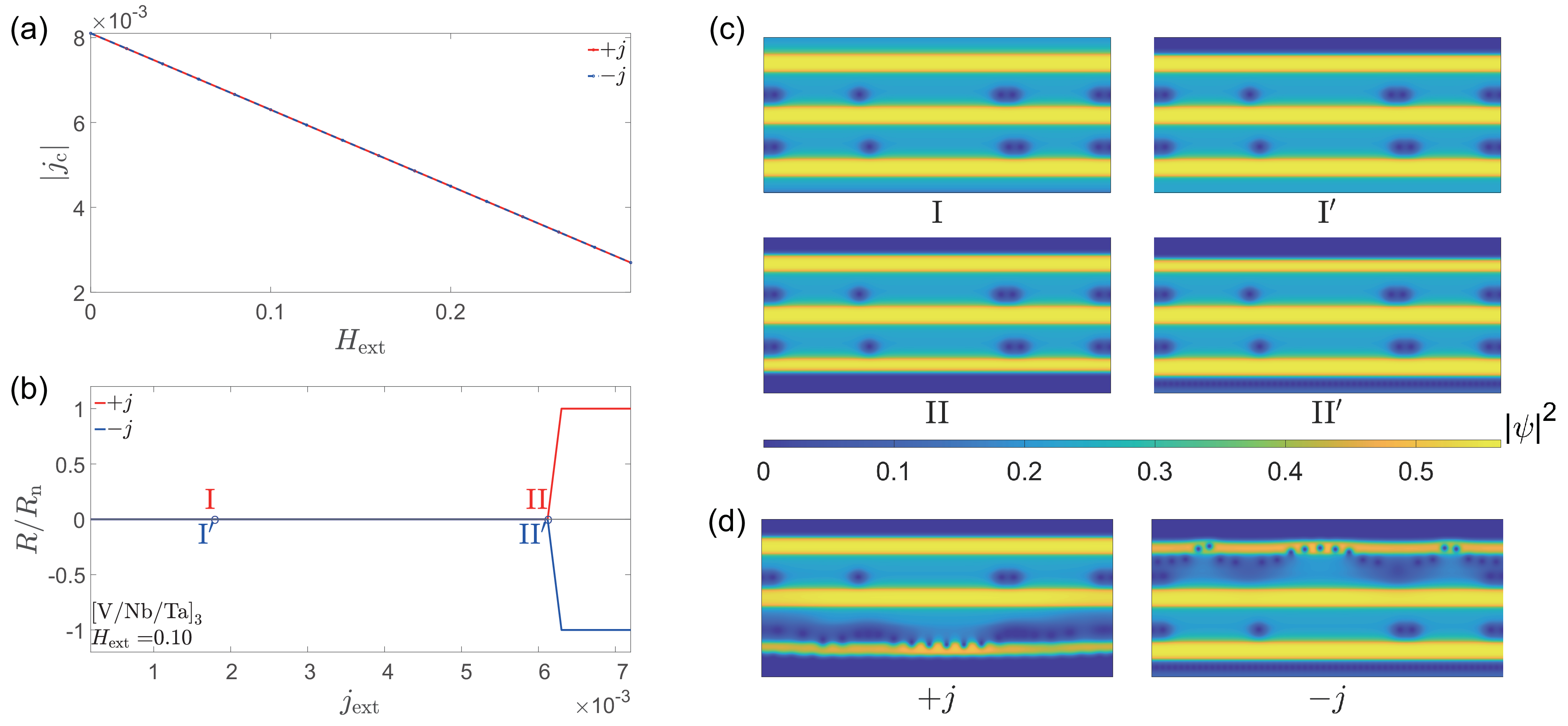}
		\caption{
			\textbf{Critical current, current–resistance behavior, and Cooper-pair density distributions in a different stacking order.}
			\textbf{(a)} Critical current as a function of the external magnetic field for the multilayer sample at a bath temperature of $T_0 = 4.0~\mathrm{K}$. The red line corresponds to the positive current direction ($+j$) and the blue line corresponds to the negative current direction ($-j$). 
			\textbf{(b)} Current–resistance characteristics under an external magnetic field of $H_\mathrm{ext} = 0.10H_\mathrm{c2}$. The resistance is normalized by the normal-state resistance $R_\mathrm{n}$. The red line corresponds to the positive current direction and the blue line to the negative current direction. Points labelled I and II on the red line and I$'$ and II$'$ on the blue line indicate representative states used in the (c), where the applied currents have equal magnitudes but opposite directions. 
			\textbf{(c)} Distributions of the Cooper-pair density for the states indicated in (b). Colors represent the magnitude of the density, with blue indicating low values and yellow indicating high values along the color bar. 
			\textbf{(d)} Cooper-pair density distributions during the phase transition for the positive and negative current directions. 
			The complete phase-transition sequence is provided in Supplementary Movie~5.
		}
		\label{fig_VNbTa}
	\end{figure*}
	
	To validate the mechanism proposed above, we examine the effect of swapping the Nb and V layers, resulting in a $\mathrm{[V/Nb/Ta]}_n$ structure. The layer thickness is set to $10\xi_0$, while the total thickness $L_z$, sample length $L_x$, and bath temperature remain unchanged. The resulting configuration is $\mathrm{[V(10\xi_0)/Nb(10\xi_0)/Ta(10\xi_0)]}_3$. Surprisingly, no SDE is observed in these superconducting heterostructures with the modified stacking order, as shown in Fig.~\ref{fig_VNbTa}(a).
	
	As a representative example, the current–resistance characteristics and CPD distributions are presented in the same manner as before, with $H_\mathrm{ext} = 0.10 H_\mathrm{c2}$. As shown in Fig.~\ref{fig_VNbTa}(b), the transition to the normal state occurs at $\lvert j_\mathrm{c}^\pm \rvert \simeq 0.0061$ for both $\pm j$. Although the flux-flow regime is nearly absent in both current directions, the SDE vanishes entirely. 
	
	The CPD distributions in Fig.~\ref{fig_VNbTa}(c) reveal that the patterns for $\pm j$ are nearly identical, aside from minor differences at the boundary layers. Furthermore, as illustrated by the complete phase-transition process in Supplementary Movie 5 and Fig.~\ref{fig_VNbTa}(d), the CPD distributions during the transition become almost symmetric for $\pm j$ (see points II and II$'$). A large number of vortices are present throughout the phase transition, and their motion generates substantial Joule heating, driving the sample into the normal state. During this process, the modified stacking order produces two competing tendencies of vortex motion: one directed from Ta to Nb and the other from V to Nb. These opposing contributions effectively counterbalance each other, leading to the disappearance of the SDE.
	
	Although vortices under $-j$ emerge earlier than those under $+j$ (points II and II$'$), the small differences in CPD and $\mathcal{F}$ between Ta and V produce only weak asymmetry, insufficient to generate an observable SDE. Consequently, $j_\mathrm{c}^\pm$  become nearly identical, indicating that the SDE is highly sensitive to the stacking sequence. This finding also suggests a simple and practical strategy for suppressing the SDE—modifying only the stacking order of the constituent layers.
	
	\section{Conclusions} 
	
	In this study, we employed the TDGL framework to investigate the SDE in asymmetric multilayer heterostructures composed of Nb, V, and Ta. Our results demonstrate that the SDE can be driven by vortex dynamics in such heterostructures under an external magnetic field, with its characteristics strongly dependent on the layer thickness.
	
	Our results indicate that when the stacking order breaks spatial inversion symmetry, as in the $\mathrm{[Nb/V/Ta]}_n$ structure, the system exhibits a pronounced SDE characterized by nonreciprocal critical currents and a direct transition to the normal state with minimal flux-flow behavior. This demonstrates excellent performance with low dissipation and underscores the importance of interlayer interactions in determining the overall superconducting response. In this regime, vortex dynamics—driven by the interplay between Lorentz forces and material-dependent preferences—play a dominant role.
	
	Furthermore, reversing the stacking order to $\mathrm{[V/Nb/Ta]}_n$ completely eliminates the SDE, indicating that the layer sequence critically determines the symmetry-breaking conditions required for the effect. The symmetry of the CPD distribution and the cancellation of opposing vortex-motion tendencies account for this disappearance. Overall, this study provides insights into the origin of the SDE in artificial superlattices and explains key experimental observations, such as those reported by Ando \textit{et al.}~\cite{ando_observation_2020}.
	
	It should be emphasized that our theoretical results are not limited to the specific $\mathrm{[Nb/V/Ta]}_n$ trilayer system. The essential requirement is a heterostructure composed of one type-I superconductor and two distinct type-II superconductors arranged in a symmetry-breaking sequence. These findings may provide insights into vortex dynamics and critical current behavior in asymmetric layered superconductors. Moreover, the demonstrated ability to eliminate the SDE through simple geometric reconfiguration offers a convenient strategy for tailoring superconducting circuit behavior in applications.
	
	%\backmatter 
	
	\section*{Acknowledgements}
	
	This work was partially supported by the National Natural Science Foundation of China under Grant No. 92565201 and the National Key R\&D Program of China under Grant No. 2024YFA1408900.
	
	\section*{Supplementary information}
	
	Supplementary information is available at the publisher's website via the DOI:
	
	https://doi.org/10.1038/s42005-025-02481-8
	
	\section*{Author Contributions}
	
	Jiong Li developed the numerical code, generated the figures, and wrote the first draft of the manuscript. 
	Ji Jiang contributed to debugging the code and discussing the results. 
	Qing-Hu Chen proposed the original idea of the study, contributed to the discussion of the results, and revised the manuscript. 
	All authors reviewed and approved the final version of the manuscript.
	
	\section*{Competing interests}
	
	The authors declare no competing interests.
	
	\section*{Data availability}
	The data generated during and/or analysed during the current study are available from the corresponding author on reasonable request.
	
	\section*{Code availability}
	The custom numerical code used in this theoretical study is available from the corresponding author on reasonable request.

	%\bibliographystyle{sn-aps}
	%\bibliography{refs}% common bib file

\end{document}